\documentclass[aps,pra, twocolumn]{revtex4}

\usepackage{graphicx}
\usepackage{amssymb}
\usepackage{amsmath}
\usepackage{colordvi}
\usepackage{color}

\newcommand{\cH}{{\cal H}}
\newcommand{\cI}{{\cal I}}

\newcommand{\tb}{\tilde{b}}

\newcommand{\vep}{\varepsilon}
\newcommand{\rev}[1]{\textcolor{black}{#1}}

\newcommand{\tcI}{\tilde{\cal I}}

\begin{document}

\title{Double-flat-top half-vortices and self-bound solitary wave billiards   in cubic-quintic media with intermodal attraction}

\author{Dmitry A. Zezyulin}

\affiliation{School of Physics and Engineering, ITMO University, St. Petersburg  197101, Russia}

\date{\today}

\begin{abstract}
	
	We consider a bimodal light field envelope propagating in a bulk medium characterized by competing cubic and quintic nonlinearities. The  subfields are coupled by a cross-phase modulation  term and experience effective attraction. We find dynamically stable stationary states which have two distinct flat-top regions with different intensities. These solutions represent half-vortices, where the first and second components are essentially different and, in particular, carry  different topological charges: zero for one component and nonzero for the other.   The typical propagation  of an  unstable half-vortex  leads to the  splitting of the central vortex core into several fragments which quasielastically interact with the boundary of the flat-top region. This behavior is interpreted as a self-bound solitary wave billiard, where the emerging fragments are the billiard balls and the flat-top region is the dynamically deforming table.  
		 
\end{abstract}

\maketitle

\section{Introduction}

A medium with cubic and quintic nonlinearities provides a basic platform for studying a competition between   nonlinear effects that come into play at different amplitudes of the propagating field.  The defocusing quintic nonlinearity stabilizes the optical beam against the collapse \cite{Fibich} and enables the formation of dynamically stable self-trapped  radially-symmetric stationary states \cite{Dimitrevski}. The  existence domain of  solitary states in the cubic-quintic (CQ) medium is constrained by an eigenvalue cutoff  \cite{Prytula}. Near this cutoff, the solutions acquire a flat-top  shape   characterized by a broad amplitude plateau. The additional quintic nonlinearity can also stabilize broad vortex states, while more narrow unstable vortices spontaneously  split into two or several fragments which are close to a fundamental (i.e., zero-vorticity) soliton  \cite{vortices}. Pairwise collisions between two stable solitons 
can follow different scenarios ranging from elastic collisions to complete coalescence.  The existence of self-bound flat-top states enables the concept of ``liquid light'', where a broad optical beam displays pressure and surface tension similar to classical liquid droplets. This analogy and its limitations have been discussed in a series of publications \cite{liquid}. In optical experiments,   competing CQ  nonlinearities have  been  realized with coherent multilevel atoms  \cite{atoms},   liquid carbon disulfide \cite{CS2}, and colloidal suspensions \cite{colloids}.  At the same time, the range of potential applications for the CQ model is broader and includes Bose-Einstein condensates with strong three-body interactions, plasma solitons, and others. From a more general perspective, competing CQ terms arise from the Taylor expansion   of any function $F(|\psi|^2)$ describing a  saturating nonlinear response. Recent collections of results on  multidimensional solitons and, in particular, vortex solitons, supported by competing nonlinearies can be found in the  review \cite{MalomedReview} and in the monograph  \cite{MalomedBook}.

Physics of competing nonlinearities becomes richer in bimodal systems, where two fields  are incoherently coupled  by a nonlinear term \cite{Maimistov99}.  There are multiple studies that report stable solutions for   bimodal fields propagating in a CQ medium. However, stable vortex solitons have been found to exist only if the the topological charges carried in each component  are equal  \cite{Mihalache2002,Mihalache2004,Desyatnikov2005} or opposite  \cite{Mihalache2004,Desyatnikov2005}. In these cases, the amplitude distributions of both fields are similar. The \textit{first goal} of this paper is to introduce a special class of solutions that form in bimodal systems characterized by competing CQ  nonlinearities and effective intermodal attraction. The amplitude profiles of these solutions  have  \textit{two}   flat-top plateaus of distinct  heights. These solutions represent half-vortices, where the first and second components are essentially different and, in particular,  have unequal propagation constants and carry  different topological charges: zero for one component and nonzero for the other. Hence the double-flat-top solutions heavily  exploit the bimodal nature of the system and have no counterparts in the single-component case. To the best of our knowledge, solutions of this type have not been discussed previously for CQ media, despite the strong interest in this topic. In addition, we show that the double-flat-top shape enhances the stability of half-vortex solutions, which lose stability as they depart  from the  double-flat-top regime.   

The \textit{second goal} of this paper is to demonstrate that the fission instability of bimodal states develops in a billiard-like behavior for emerging solitary waves. In this scenario, two (or more) small fragments resulting from the splitting of an unstable vortex represent the billiard balls, and  the flat-top plateau   works as a dynamically deforming table that confines the billiard within a bounded domain. Therefore, in contrast to previously studied solitary-wave billiards \cite{Prati10,Prati,Cuevas}  constrained by external boundaries, we present  a self-bound solitary-wave billiard.

The paper is organized as follows. In Sec.~\ref{sec:model}, we discuss the two-component model and develop a simple asymptotic expansion which describes the bifurcation of solutions from a certain asymptotic limit. In Sec.~\ref{sec:num},  we present the main numerical results on  the existence and stability of double-flat-top solutions, as well as on  the  billiard-like dynamics. Section~\ref{sec:concl} provides a conclusion.
 
\section{The model and analytical results} 

\label{sec:model}

We consider propagation of a two-component envelope $(\psi_1, \psi_2)$   described by a pair of equations that  can be compactly presented in the form
\begin{equation}
\label{eq:main}
i\frac{\partial \psi_{1,2}}{\partial z}  +\frac{1}{2} \left(\frac{\partial^2 \psi_{1,2} }{\partial x^2} + \frac{\partial^2  \psi_{1,2}}{\partial y^2}\right)    -  \frac{\partial \cH(\cI_1, \cI_2)}{\partial \cI_{1,2} } \psi_{1,2} = 0,
\end{equation}
where $z$ is the longitudinal coordinate, $(x,y)$ are the transverse coordinates,    $\cH(\cI_1, \cI_2)$ is the Hamiltonian density corresponding to the nonlinear self-action and coupling, and $\cI_{1,2} = |\psi_{1,2}|^2$.  A prototypical example which will be used below for presentation of the main results corresponds to
\begin{eqnarray}
\label{eq:Hamilt}
\cH = -\frac{(\cI_1^2 +  \cI_2^2)}{2}  + \frac{(\cI_1^3  + \cI_2^3)}{3}  \nonumber  \\- \beta \cI_1 \cI_2  + \alpha \cI_1\cI_2(\cI_1     + \cI_2).
\end{eqnarray}

The model defined by Eq.~(\ref{eq:main}) and  the Hamiltonian density (\ref{eq:Hamilt}) was introduced in Ref.~\cite{Maimistov99} to study the attraction between multidimensional solitons in a bimodal system, where  two   waves carry   orthogonal polarizations or have  different carrier wavelengths and are incoherently coupled via  cross-phase modulation (XPM).  Later, this model and its modifications have been intensively discussed in the literature \cite{Mihalache2002,Mihalache2004,Desyatnikov2005, misc}. Coefficient $\beta$ represents the cubic   XPM, and $\alpha$ is the   quintic self-defocusing XPM coefficient.  \rev{For two pulses propagating with distinct carrier frequencies, the XMP coefficient is $\beta=2$. For two field with the same   frequency and different polatizations, $\beta=2/3$.  More generally, $\beta$ can vary from 2/3 to 2 in elliptically birefringent fibers \cite{Agrawal}.}

We introduce substitution for stationary states  $\psi_{1,2} = e^{i b_{1,2} z} u_{1,2}(x,y)$, where $b_{1,2}$ are real propagation constants, which are generically different. The stationary equations take the form
\begin{equation}
\label{eq:main:stat}
\frac{1}{2} \Delta u_{1,2}   - b_{1,2} u_{1,2}  -    \left.\frac{\partial \cH(\cI_1, \cI_2)}{\partial \cI_{1,2}}\right|_{\cI_{1,2} = |u_{1,2}|^2}  u_{1,2}=0,
\end{equation}
where $\Delta = \partial^2/\partial x^2 + \partial^2/\partial y^2$. We consider solutions with   different amplitude distributions in the two components. In the first component, the intensity corresponds to a (generically vortical)  solitary wave of relatively small width. The second component has zero topological charge and is composed of a solitary wave sitting on top of a wider plateau.    The existence of solutions of this type  can be anticipated from an asymptotic procedure outlined below.   It starts from the limit where the intensity of the first component is identically zero,   $u_1\equiv 0$. The propagation constant of the second component corresponds to the cutoff value, $b_2 := \tb_2$,  and the  intensity of the second component is spatially uniform, $\cI_2 := \tcI_2 = \mathrm{const}$. This  intensity   can be determined from the condition
\begin{equation}
\label{eq:1}
\frac{\partial \cH(0, \tcI_2)}{\partial \cI_2}  \tcI_2 =  \cH(0,  \tcI_2),
\end{equation} 
and the cutoff  propagation constant, $\tb_2$, can be computed  as 
\begin{equation} 
\label{eq:tmu2}
\tb_2 :=  -{\partial \cH(0, \tcI_2)} / {\partial \cI_2}.
\end{equation}
Departing from this  limit, we look for  solutions with $\cI_1 \ll 1$ and $|\cI_2 - \tcI_2| \ll 1$ using   asymptotic expansions with respect to a certain small parameter $\vep \ll 1$. The solutions can be constructed  in the form of expansions  
\begin{eqnarray}
\label{eq:u1}
u_1 &=& \vep p_1 + \vep^3 p_3 + \ldots,\\[1mm] 
\label{eq:u2}
u_2 &=& {\tcI_2}^{1/2}(1+\vep^2 q_2 + \vep^4 q_4+ \ldots ).
\end{eqnarray}
The functions $p_j$ corresponding to the first component are generically complex-valued, while the functions $q_j$ corresponding to the second component are real-valued. The unknown functions $p_j=p_j(X,Y)$ and $q_j = q_j(X,Y)$ depend on  the slow-scale transverse   variables, $X = \vep x$, $Y = \vep y$,   and satisfy   zero boundary conditions as $X^2 + Y^2 \to \infty$. The small parameter $\vep$ characterizes the deviation of the propagation constant of the first component from the asymptotic  limit: $\vep^2 = b_1 -\tb_1$,
where  $\tb_1$ is readily determined from the leading order of the perturbation procedure:
\begin{equation} 
\label{eq:tmu1}
\tb_1 :=  -{\partial \cH(0, \tcI_2)} / {\partial \cI_1}.
\end{equation}
The propagation constant of the second component remains fixed at this moment due to the uniform density at infinity: hence $b_2 = \tb_2$.  

In the next order of the expansions, we obtain
\begin{equation}
\label{eq:cubic}
\frac{1}{2} \left (\frac{\partial^2 p_1}{\partial X^2} + \frac{\partial^2 p_1}{\partial Y^2}\right )  - p_1 + \sigma |p_1|^2p_1 = 0, 
\end{equation}
where $ \sigma = -(h_{20} h_{02} - h_{11}^2)/{h_{02}}$ and, for compactness, we use the following notation for partial derivatives of the Hamitonian density:
\begin{equation}
\label{eq:pd}
h_{jk} = \frac{\partial^{j+k} \cH(0, \tcI_2)}{\partial \cI_1^j \partial \cI_2^k}.
\end{equation}
In addition, we obtain  $q_2 = -h_{11} |p_1|^2 / (2{\tcI_2} h_{02})$.

Equation (\ref{eq:cubic}) is   a stationary version of the nonlinear Schr\"odinger equation with cubic nonlinearity. If $\sigma>0$, then the cubic term is self-focusing, and this equation has solitary wave solutions which vanish as $X^2+Y^2\to \infty$ and  carry an integer  topological charge. If $h_{11}<0$  and $h_{02}>0$ (which is the case considered below), then the function $q_2$ is positive, which means than the effective intermodal attraction produces a localized increase of intensity in the second component. This high-intensity region is located on top  of the uniform background  given by  $\tcI_2$.  In addition, if the propagation constant of the second component is chosen slightly below  the cutoff value, i.e., $b_2 = \tb_2 - \delta$, where $1 \gg \delta >0$,  then the infinite background transforms to a flat-top profile of finite transverse extent, and both components $u_1$ and $u_2$ satisfy zero boundary conditions as $x^2 + y^2 \to 0$. 

\begin{figure}
	\begin{center}
		\includegraphics[width=0.999\columnwidth]{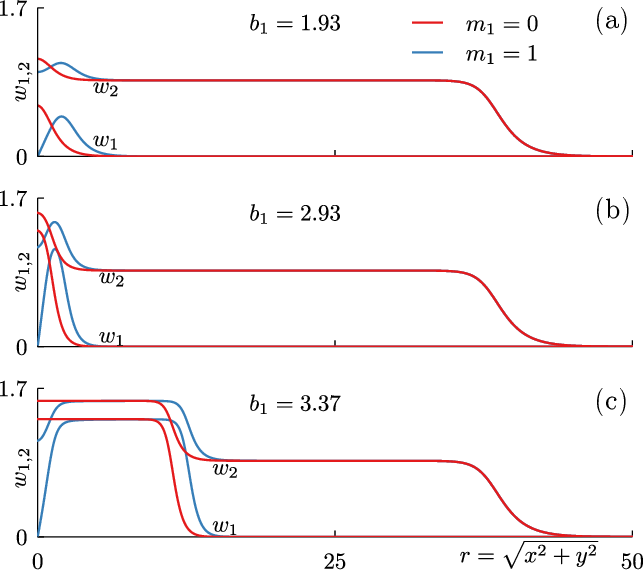} 
	\end{center}
	\caption{Radial profiles of stationary states with zero ($m_1=0$) and single ($m_1=1$) topological charge in the first component for three different  propagation constants  $b_1$ in the first component and fixed propagation constant in the second component, $b_2 = 0.1835$ (which is slightly below the cutoff propagation constant $\tb_2 = 3/16\approx 0.1875$). The lowest panel illustrates the double-flat-top solutions. The topological charge of the second component is always zero: $m_2=0$. }
	\label{fig:profiles}
\end{figure}

\section{Numerical results} 
\label{sec:num}

\subsection{Stationary solutions and their stability}

The solutions predicted by the asymptotic procedure above can be used as a starting point for numerical continuation. To present the numerical results, we will consider the Hamiltonian density (\ref{eq:Hamilt}), where the coefficient of effective intermodal attraction  is chosen as   $\beta=2$, which corresponds to  two carrier frequencies or two circular polarizations \cite{Maimistov99,Agrawal}. Then   $h_{11} = -2 + 3\alpha /2$, $h_{02} = 1/2$, and $\sigma = 9(1-\alpha)(2-\alpha)/2$.  If  $\alpha$ is zero or small, then  $\sigma>0$, and Eq.~(\ref{eq:cubic}) has localized solitary-wave solutions.  Below  we use the simplest (``minimal'') model with  $\alpha=0$ \cite{Mihalache2004}. The results will not be affected significantly for small nonzero $\alpha$. However, the case of large $\alpha$  should be considered separately.

\begin{figure}
	\begin{center}
		\includegraphics[width=0.999\columnwidth]{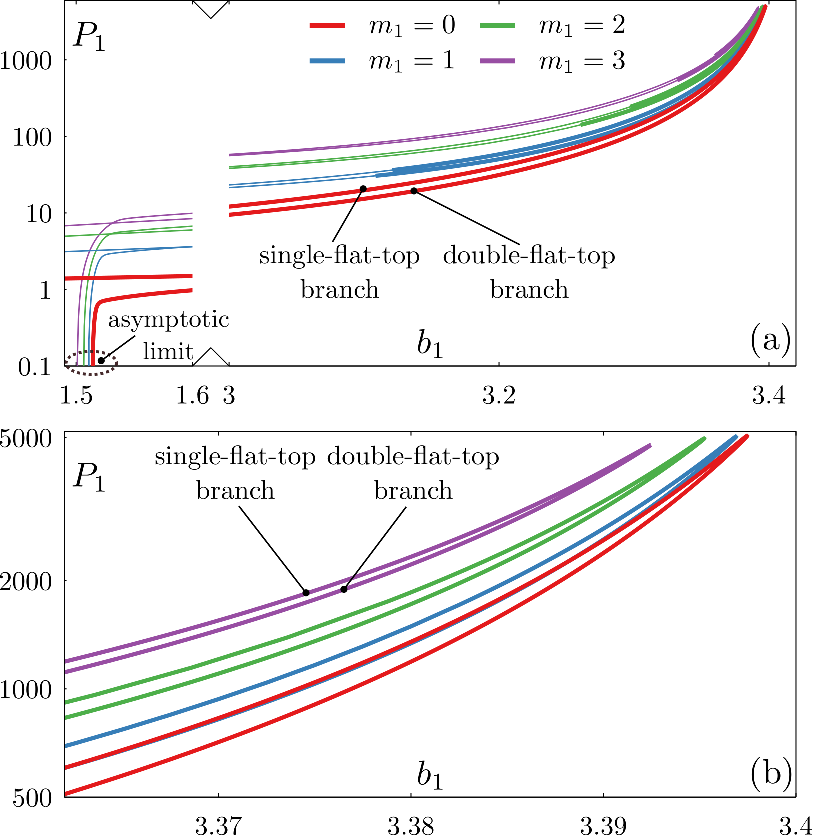} 
	\end{center}
	\caption{(a) Dependencies of the energy flow in the first component, $P_1$, on the propagation constant in the same component, $b_1$, for several values of the topological charge, $m_1=0,1,2,3$, while the propagation constant in the second component, $b_2$, is fixed to the same value as in Fig.~\ref{fig:profiles}. Fragments with thin and thick lines represent unstable and stable solutions, respectively. To improve the clarity, we have omitted the interval $b_1 \in (1.6, 3)$ from the horizontal axis. All solutions in the omitted  range are unstable, except for those with $m_1=0$. The dashed ellipse with the label ``asymptotic limit'' highlights the region where the solutions are born in the asymptotic limit, i.e., as   $\vep \to 0$.  Panel (b) shows a closer view of the same dependencies, focusing on the region where the double-flat-top branches meet with the single-flat-top ones. }
	\label{fig:families}
\end{figure}

In Fig.~\ref{fig:profiles} we illustrate  the transformation  of solutions with zero ($m_1=0$) and single ($m_1=1$) topological charge in the first component  for three different values of  the   propagation constant of the first component, i.e., $b_1$. The second propagation constant, $b_2$,  is kept fixed slightly below the cutoff value $\tb_2$.  As usual, the radially-symmetric solutions are represented as $u_{1,2} = w_{1,2}(r)e^{im_{1,2} \varphi}$, where $r$ and $\varphi$ are the polar coordinates  and  $m_1$ and $m_2$ are integer topological charges for the first and second components, respectively. In this study, we consider solutions with $m_2=0$. The solutions are born at $b=\tb_1 = 3/2$,  see the dashed ellipse in Fig.~\ref{fig:families}(a).  For $b_1$ close to  $\tb_1$, the first component is small, and  the second component has a small-amplitude solitary wave located on top of a broad plateau [Fig.~\ref{fig:profiles}(a)].  As one moves away from the asymptotic limit, the amplitude of this solitary wave increases [Fig.~\ref{fig:profiles}(b)]. Most interestingly, for a sufficiently large $b_1$ [Fig.~\ref{fig:profiles}(c)], this solitary wave   broadens dramatically, forming a secondary plateau atop the already existing  one. This results in the emergence of a double-flat-top solution.

\begin{figure}
	\begin{center}
		\includegraphics[width=0.999\columnwidth]{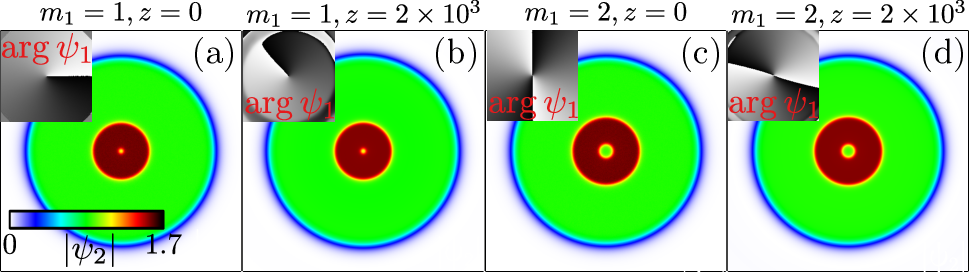}
	\end{center}
	\caption{Examples of stable propagation of double-flat-top states with vorticity $m_1=1$ (a,b) and $m_1 = 2$ (c,d). Large pseudocolor plots show the amplitude distribution, $|\psi_2|$, in the  \emph{second} component at $z=0$ (a,c) and $z=2\times 10^3$ (b,d). The insets show the pseudocolor plots of the phase distribution, $\varphi = \arg \psi_1$, in the \emph{first} component, where black and white  colors correspond to $\varphi=-\pi$ and $\varphi=\pi$, respectively. Large plots show the amplitude distributions within the window $(x,y) \in [-50, 50]\times [-50,50]$, and the insets show the phase distributions  within the window $(x,y) \in [-15, 15]\times [-15,15]$. The shown solutions correspond to the propagation constant $b_1 =  3.3674$. }
	\label{fig:stable}
\end{figure}

The diameter of the high-intensity plateau increases with the increase of $b_1$, while the diameter of the low-intensity plateau is fixed by the second propagation constant $b_2$. As a result, for   large $b_1$, both plateaus are of approximately the same size, and the shape of solution  becomes single-flat-top.   In Fig.~\ref{fig:families}(a) which presents the families of solutions as dependencies of the energy flow in the first component, $P_1 = \iint_{\mathbb{R}^2} |u_1|^2dxdy$, on the propagation constant $b_1$, each branch of double-flat-top solutions eventually makes a U-turn merging with a corresponding branch of single-flat-top solutions. The region where the double-flat-top branches meet with the single-flat-top ones  is additionally magnified  in Fig.~\ref{fig:families}(b). 

The  expansions developed above predict that, close to the bifurcation from the asymptotic limit,  the  solutions are described by the cubic nonlinear Schr\"odinger equation. The solutions of the latter equation are unstable with respect to collapse \cite{Fibich}. In the CQ  medium the collapse is suppressed. However, the question of the stability of the found solutions remains a significant  concern. We have examined  the linear stability  using the standard substitution for   perturbed solutions in the form $\psi_{1,2} = e^{i b_{1,2} z} [w_{1,2}(r) e^{im_{1,2} \varphi} + \sum_{k=0}^\infty A_{1,2}^{(k)}(r)e^{\lambda z + i(m_{1,2} + k)\varphi} + B_{1,2}^{(k)}(r)e^{\lambda^* z + i(m_{1,2} - k)\varphi} ]$.  We have considered solutions with topological charges  $m_1=0,1,2,3$ and $m_2=0$. Further, the functions  $A_{1,2}^{(k)}(r)$  and $B_{1,2}^{(k)}(r)$  represent the  radial profiles of small-amplitude perturbations, and integer $k$ is the azimuthal index of the perturbation. Complex $\lambda$ determines the behavior of the perturbation along the propagation distance. A solution is unstable if the real part of $\lambda$ is positive. Using the perturbed solution in Eqs.~(\ref{eq:main}) and performing the linearization procedure with respect to  $A_{1,2}$ and $B_{1,2}$, for each $k$  we obtain an   eigenproblem    for the eigenvalue $\lambda$ (the eigenproblems with different $k$ decouple). 

We have numerically solved the linear stability eigenvalue problems for perturbations with  azimuthal indexes ranging from $k=0$ to $k=10$. The results indicate that the zero-charged solutions with $m_1=m_2=0$ are stable. The solutions with nonzero $m_1$ are unstable right after  the bifurcation from the asymptotic limit. However, as the propagation constant $b_1$ increases and the solutions develop the double-flat-top shape, the instability increment vanishes.  Hence, even for a nonzero vorticity $m_1\ne 0$ the double-flat-top solutions are stable. In Fig.~\ref{fig:families} we highlight the parts of the curves corresponding to stable solutions using  thick lines. Soon after the U-turn, which connects the double-flat-top and single-flat-top branches, the vortex solutions become unstable again.

To confirm the stability of double-flat-top half-vortices, we have simulated nonlinear propagation of corresponding solutions integrating Eq.~(\ref{eq:main}) with a split-step method. In each case, a small amount of random noise with an amplitude of a few percent of the original solution was added to the input fields $\psi_{1,2}$. In Fig.~\ref{fig:stable} we show stable propagation of  double-flattop states with single ($m_1=1$) and double ($m_1=2$) topological charge in the first component. The solutions preserve their amplitude distributions and phase structures for indefinitely long propagation distances.

\begin{figure*}
	\begin{center}
		\includegraphics[width=0.999\textwidth]{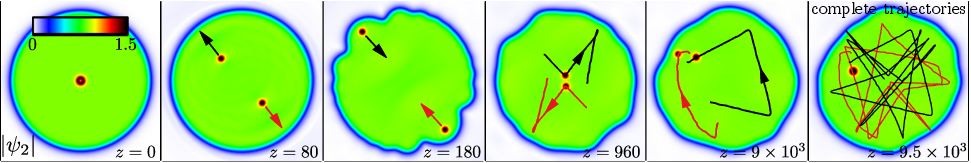}%
	\end{center}
	\caption{Propagation of an unstable state with the unitary topological charge in the first component, i.e., $m_1=1$ and with $b_1 = 3.03$. The snapshots taken at different propagation distances $z$ show the amplitude  of the second component, $|\psi_2|$, within the window $(x,y) \in [-45, 45]\times [-45, 45]$. The lines  with arrows show the fragments of trajectories along which the emerged fragments move, and the direction of the arrows indicates an increase in the propagation distance $z$. The rightmost plot shows the amplitude distribution at the propagation distance past the irreversible coalescence of two fragments into a single one. In this plot, the complete trajectories of both fragments are shown, starting from the propagation distance where they have emerged from the vortex splitting instability and ending at the propagation distance where they have coalesced.  The amplitude of the first component, $|\psi_1|$, behaves in a similar way, but the high-intensity spots move on a zero background, i.e., there is no flat-top plateau in the first component. See Supplemental Material for the multimedia file  corresponding to this simulation.}
	\label{fig:m1unstable}
\end{figure*}

\begin{figure}
	\begin{center}
			\includegraphics[width=0.999\columnwidth]{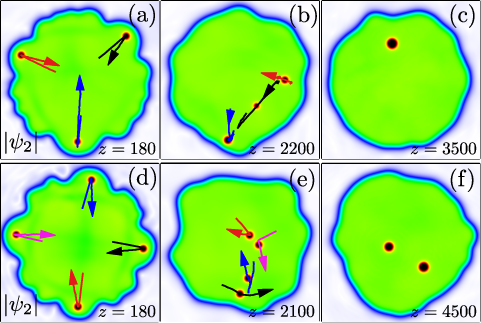}%
	\end{center}
	\caption{Examples of unstable propagation for  states with double ($m_1=2$, a--c) and triple ($m_1=3$, d--f)   topological charge in the first component.  The   snapshots  taken at different propagation distances $z$ show the amplitude of the second component, $|\psi_2|$, within the window $(x,y) \in [-45, 45]\times [-45, 45]$. The lines  with arrows show the fragments of trajectories along which the emerged fragments move, and the direction of the arrows indicates an increase in the propagation distance $z$.  The propagation constants corresponding to  the first components are equal to $b_1 \approx 3.01$ (a--c) and  $b_1 \approx 3.03$ (d--f). See Supplemental Material for the
		multimedia files corresponding to these simulations.}
	\label{fig:m23unstable}
\end{figure}

Away from the double-flat-top regime, the states with nonzero vorticity in the first component are unstable. Figures~\ref{fig:m1unstable} and \ref{fig:m23unstable}  illustrate the propagation of solutions which  contain a narrow vortex core embedded within a  broad flat-top region, similar to those  shown in Fig.~\ref{fig:profiles}(b). During the propagation of an unstable solution, the   central vortex core splits into several high-intensity spots of approximately equal  size.  These fragments move away from the center and quasielastically reflect off the boundary of the broad flat-top region. Hence the dynamics is confined within a bounded domain. This behavior contrasts sharply with the typical instability of vortex states in CQ media, where the emerging  fragments fly indefinitely away from each other.

Figure~\ref{fig:m1unstable} shows the behavior the   unstable state with a unitary charge in the first component, i.e., with $m_1=1$. The input solution has been  perturbed with a weak random noise. The central core spontaneously splits into two  high-intensity spots,  which quasielastically reflect off the boundary of the flat-top plateau and return to the center of the flat-top region. There, they undergo a quasielastic collision and then move towards the boundary again. During the subsequent propagation, these two fragments experience multiple reflections off the boundary and,  occasionally, several mutual collisions. The specific scenario of interaction between the two spots depends on their relative phase computed in the first component. If the phase difference is $\pm \pi$, then the collision is nearly perfectly elastic, and if the phase difference is close to  zero, then the fragments coalesce irreversibly. Intermediate values of the phase difference result in quasielastic collisions, accompanied by a  redistribution of the field. As a result of this interaction, one of the spots increases in size, while the other decreases.  Due to the universal  phase loss  which is ubiquitous in nonlinear nonintegrable systems \cite{loss}, after a long propagation distance the phase difference between the fragments can be arbitrary. Therefore, in the limit $z\to \infty$ the fragments are expected to coalesce, which is indeed observed for sufficiently large propagation distances, see the rightmost plot in Fig.~\ref{fig:m1unstable}.

Figure~\ref{fig:m23unstable}  shows the examples of  propagation of unstable states with double [$m_1=2$, Fig.~\ref{fig:m23unstable}(a--c)] and triple [$m_1=3$, Fig.~\ref{fig:m23unstable}(d--f)] topological charge in the first component. The number of fragments that emerge  from the splitting of an unstable vortex core  is, in general, different, and depends on the topological charge $m_1$ and on the particular realization of the random noise applied to the input beam. In particular, for the unstable solution with $m_1=3$ we have encountered the situations where the central core breaks into three or four parts [the latter case is shown in  Fig.~\ref{fig:m23unstable}(d--f)]. The emerging fragments remain confined within a bounded domain and undergo both quasielastic and inelastic collisions. The latter result in coalescence.

Adopting the  liquid light analogy \cite{liquid}, the unstable solutions evolve  into a dilute  emulsion of several  high-density droplets dispersed within a low-density liquid. The motion of the dispersed phase is constrained by the surface tension at the interface of the larger droplet. Interactions between the dispersed droplets can be quasielastic or irreversibly cohesive. Consequently, as the propagation distance increases, the dispersed droplets coalesce into a single blob that floats  within the larger, lower-density droplet.

\subsection{Billiard-like dynamics}

The transient patterns that emerging during the propagation of the initially unstable states in Figs.~\ref{fig:m1unstable} and  \ref{fig:m23unstable} can be interpreted as self-bound solitary wave billiards. The high-intensity fragments correspond to the billiard balls, and the wide flattop region acts as a  deforming billiard table.  Typically, the concept of a billiard  involves a particle or wave interacting with a closed boundary  whose shape is either fixed or driven from the outside. For wave billiards, including  a recently proposed concept of a solitary-wave billiard \cite{Cuevas}, the  walls are usually modelled by an  external boundary conditions (say the Dirichlet one). In an experimental realization of a cavity soliton billiard \cite{Prati}, the dynamics was bounded by a spatially structured external pump. In our case, the billiard-like dynamics are confined by the solution itself. Moreover, the boundary of the billiard table deforms dynamically due to its interaction with the ball(s).

\begin{figure}
	\begin{center}
			\includegraphics[width=0.999\columnwidth]{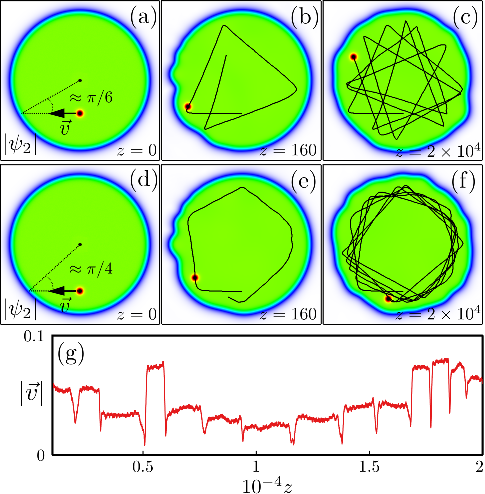}
	\end{center}
	\caption{Panels (a--c) and (d--f) show two examples of self-bound single-ball billiards which correspond to two different position of the ball at $z=0$ and the same initial velocity  indicated at (a,d); panels (b,e) show the amplitude distributions at the propagation distance past  the first collision between the ball and the boundary and the initial parts of the travelled trajectories; panels  (c,f) show the trajectories travelled at $z=2\times 10^4$.   The snapshots  taken at different propagation distances $z$ show the amplitude of the second component, $|\psi_2|$, within the window $(x-x_0,y-y_0) \in [-45, 45]\times [-45, 45]$, where $(x_0, y_0)$ is the approximate center of the billiard. Panel (g) shows the instantaneous ball velocity for the billiard shown in (a--c) within the interval $z\in [10^3, 2\times 10^4]$. See Supplemental Material for the 		multimedia files corresponding to these simulations.}
	\label{fig:billiards}
\end{figure}

In Figs.~\ref{fig:m1unstable} and \ref{fig:m23unstable}  the billiard-like dynamics  emerge spontaneously. One can also create an on-demand billiard using a fundamental (zero-vorticity) solution as a ball and imparting an initial velocity via a phase tilt  of the first components' wavefunction, $\psi_1 \to \psi_1 e^{i(k_x x + k_y y)}$, where  $k_x$ and $k_y$ determine the velocity components. This phase tilt is performed only  in a localized spatial region around the higher-intensity spot and only for the first component. Two examples of the resulting dynamics are shown in Fig.~\ref{fig:billiards}. They correspond to different angles of incidence between the initial velocity vector and the circular boundary of the flat-top region.  The billiard-like  dynamics is apparently chaotic and  persists for an indefinitely large propagation distance. \rev{As a part of the    ball's initial velocity  is transferred to the boundary of the flat-top region, it causes a slow and nearly uniform drift of the entire billiard  in the corresponding direction. Consequently, to clearly display the ball's trajectory $(x_b(z), y_b(z))$ over a large propagation distance in Fig.~\ref{fig:billiards}(c) and (f), we subtract this systematic drift using the transformation:
$x_b \to x_b -  v_x  z$ and $y_b \to y_b -   v_y z $, where    $v_x$ and $v_y$ are   the slow drift velocities   in the horizontal and vertical directions, respectively. For the numerical results shown in Fig.~\ref{fig:billiards}, $v_x\sim 10^{-3}$ and $v_y \sim 10^{-6}$.}

The first collision between the ball and the boundary causes a significant reduction in the velocity of the reflected ball. The energy lost by the ball excites the surface waves that propagate along  the interface  of the light droplet.  Subsequent collisions with the boundary lead to only minor further changes in the ball's speed.  In Fig.~\ref{fig:billiards}(g) we plot the instantaneous velocity of the ball  within the  interval $z\in [10^3, 2\times 10^4]$. It should be noticed that a collision  with the boundary can also accelerate the ball.

\rev{The nearly elastic reflection of a high-intensity fragment from the boundary of the   flat-top region can be viewed as a form of nonlinear total internal reflection. In this analogy, a narrow beam is reflected at the interface of a high-refractive-index domain created by the broad flat-top soliton. All our results pertain to narrow beams incident at sufficiently small angles, and we have never observed appreciable refraction of a beam out of the flat-top region. }

\begin{figure}
	\begin{center}
		\includegraphics[width=0.999\columnwidth]{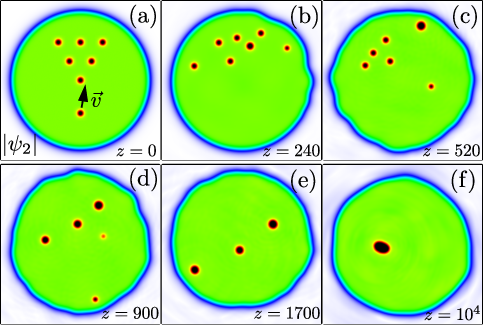}
	\end{center}
	\caption{\rev{A multiball billiard, where a ball with   initial velocity $\vec{v}$ breaks a pyramid-shaped group of several   stationary balls.   See Supplemental Material for the 		multimedia file  corresponding to this simulation.}}
	\label{fig:pyramid}
\end{figure}

\rev{Figure~\ref{fig:pyramid} presents an example of a multiball billiard. The input field is prepared as a combination of a single ball incident on a  pyramid-shaped group of six stationary balls. The initial phases of all balls  were chosen randomly from the interval $[-\pi, \pi)$. The fragments undergo multiple  quasielastic and inelastic collisions and remain confined within the flat-top region.   At a sufficiently large propagation distance, they   coalesce  into a single fragment.}

\section{Discussion and conclusion}
\label{sec:concl}

This paper has focused on two   results regarding the propagation of a bimodal light field in a medium with cubic-quintic nonlinearity and effective attraction between the components. The first result consists in the existence of dynamically stable  double-flat-top states  carrying nonzero vorcitity in one of the subfields. These states form when the propagation constant of one component is brought close to the cutoff, and  sufficiently strong intermodal attraction creates a secondary flat-top plateau atop  the already existing one.  The flat-top phase is a characteristic property of nonlinear states that arise under the action of competing nonlinearities, near the eigenvalue cutoff.  Therefore, this finding represents an important   contribution to the existing body of knowledge. In particular, it can  enrich and deepen the analogy between the propagation of a light beam and the behavior of classic liquids. Very recently,  two-dimensional double-flat-top solutions (without any vorticity) have been found in atomic mixtures \cite{KZ25}, where the competition between intra- and intercomponent interactions leads to formation of liquidlike quantum droplets \cite{Petrov2016}.  However, the effective nonlinearities that describe the formation of quantum droplets  have rather special  forms which, in addition, depend on the effective dimensionality of the  atomic condensate  \cite{Petrov2015,Petrov2016}. The results of this study are based on the paradigmatic cubic-quintic model and provide strong evidence that vortical double-flat-top states are ubiquitous and can be realized in a wide range of bimodal systems with competing nonlinearities. \rev{While the double flat-top states have been computed numerically, they are also expected to be accessible via a variational approach for vortex solitons in cubic-quintic media \cite{Paredes22}.}

The second result of this paper is   the billiard-like dynamics,   where high-intensity fragments interact quasi-elastically with the   boundary   of the flat-top region. Consequently, this solitary-wave billiard is self-bound, requiring no external confinement. Collisions  between the billiard-ball fragments can range from nearly elastic reflection to irreversible coalescence.   A single-ball billiard remains self-confined and persists over indefinitely large propagation distances. The billiard orbits are open, and the dynamics  appear to be chaotic.  In this respect, the observed behavior can provide an example of a self-localized  spatiotemporal chaotic state, similar to chaoticons \cite{chaoticon,vortexchaoticon}.

\section*{Acknowledgments}
The research (asymptotic and numerical analysis, interpretation of the results and writing of the paper) was supported by Russian Science Foundation, Grant No. 25-12-00119 \cite{RNF}. The research was supported  by Priority 2030 Federal Academic Leadership Program.

\section*{Data availability}

The data that support the findings of this article are openly available \cite{data}.



\end{document}